\documentclass[iop,apjl,12pt]{emulateapj}
\usepackage{times}
\usepackage[normalem]{ulem}
\usepackage{apjfonts}
\usepackage{epsfig}
\usepackage{natbib}
\usepackage{amsfonts}
\usepackage{amsmath}
\usepackage{multirow}
\usepackage{enumerate}
\usepackage{placeins}
\usepackage{enumitem}
\usepackage{float}
\usepackage[usenames]{xcolor}
\usepackage[plainpages=false, colorlinks=true, anchorcolor=lblui, linkcolor=blue, citecolor=black, urlcolor=blue, bookmarks=false]{hyperref}
\sloppypar
\newcommand\cqg{{Class. Quantum Grav.}}

\def\Dwa{$\,$\uppercase\expandafter{\romannumeral5}$\,$}

\def\sless{\lower2pt\hbox{$\buildrel {\scriptstyle <}
   \over {\scriptstyle\sim}$}}

\def\sgreat{\lower2pt\hbox{$\buildrel {\scriptstyle >}
   \over {\scriptstyle\sim}$}}
\def\sharpnull#1{}

\newcommand{\code}[1]{\texttt{#1}}

\newcommand{\shortauth}{Ott \emph{et al.}}
\newcommand{\slugcom}{Submitted on 2017 December 4}
\slugcomment{\slugcom}

\lefthead{\sc \footnotesize \hspace*{1em} \slugcom \hfill \shortauth\hspace*{0.2em}}
\righthead{\sc \footnotesize \hspace*{1em} \slugcom \hfill \shortauth\hspace*{0.2em}}

\begin{document}

\title{The Progenitor Dependence of Three-Dimensional Core-Collapse Supernovae}
\author{Christian D. Ott\altaffilmark{1,2}}
\author{Luke F. Roberts\altaffilmark{3}}
\author{Andr\'e da Silva Schneider\altaffilmark{2}}
\author{Joseph M.~Fedrow\altaffilmark{1}}
\author{Roland Haas\altaffilmark{4}}
\author{\\ Erik Schnetter\altaffilmark{5,6,7}}
\altaffiltext{1}{Yukawa Institute for Theoretical Physics, Kyoto University, Kyoto, Japan}
\altaffiltext{2}{TAPIR, Mailcode 350-17, California Institute of Technology, Pasadena, CA 91125, USA, christian.d.ott@gmail.com}
\altaffiltext{3}{National Superconducting Cyclotron Laboratory, Michigan State University, East Lansing, MI 48824, USA}
\altaffiltext{4}{National Center for Supercomputing Applications, University of Illinois, Urbana, IL 61801, USA}
\altaffiltext{5}{Perimeter Institute for Theoretical Physics, Waterloo, ON, Canada}
\altaffiltext{6}{Department of Physics, University of Guelph, Guelph,
  ON, Canada}
\altaffiltext{7}{Center for Computation \& Technology, Louisiana State
  University, Baton Rouge, LA, USA}

\begin{abstract}
We present a first study of the progenitor star dependence of the
three-dimensional (3D) neutrino mechanism of core-collapse
supernovae. We employ full 3D general-relativistic multi-group
neutrino radiation-hydrodynamics and simulate the post-bounce
evolutions of progenitors with zero-age main sequence masses of $12$,
$15$, $20$, $27$, and $40\,M_\odot$. All progenitors, with the
exception of the $12\,M_\odot$ star, experience shock runaway by the
end of their simulations. In most cases, a strongly asymmetric
explosion will result. We find three qualitatively distinct evolutions
that suggest a complex dependence of explosion dynamics on progenitor
density structure, neutrino heating, and 3D flow. (1) Progenitors with
massive cores, shallow density profiles, and high post-core-bounce
accretion rates experience very strong neutrino heating and
neutrino-driven turbulent convection, leading to early shock
runaway. Accretion continues at a high rate, likely leading to black
hole formation. (2) Intermediate progenitors experience
neutrino-driven, turbulence-aided explosions triggered by the arrival
of density discontinuities at the shock. These occur typically at the
silicon/silicon-oxygen shell boundary. (3) Progenitors with small
cores and density profiles without strong discontinuities experience
shock recession and develop the 3D standing-accretion shock
instability (SASI). Shock runaway ensues late, once declining
accretion rate, SASI, and neutrino-driven convection create favorable
conditions.  These differences in explosion times and dynamics result
in a non-monotonic relationship between progenitor and compact remnant
mass.
\end{abstract}

\section{Introduction}

Core-collapse supernovae (CCSNe) are the birth places of neutron stars
and black holes. They liberate the ashes of stellar evolution, seeding
the interstellar gas with the elements from which planets form and
life is made. They feed back on star formation and regulate galaxy gas
budgets. Yet, despite their importance for much of astrophysics, our
understanding of the CCSN explosion mechanism, and its
dependence on progenitor star properties, is woefully incomplete.

The CCSN problem (see, e.g., \citealt{janka:07} for an in-depth
review) boils down to the total pressure behind the shock having to
offset the accretion ram pressure of the outer core impinging on the
shock. The hot accreting protoneutron star (PNS) formed at core bounce
emits a huge flux of neutrinos of all species. The \emph{neutrino
  mechanism} \citep{bethe:85} relies on a fraction of these neutrinos
being reabsorbed in a ``gain layer'' below the shock. There, they heat
the gas, increasing the thermal pressure. Stars are spherical from a
distance and much of the early CCSN simulation work was conducted in
spherical symmetry (1D). But 1D simulations fail to show explosions
powered by the neutrino mechanism for all but the lowest-mass
progenitors ($M\lesssim10\,M_\odot$; e.g., \citealt{kitaura:06}).
Neutrino heating is strongest at the base of the gain layer and it
establishes a radially decreasing gradient in entropy. The first
axisymmetric (2D) simulations showed that turbulent convection driven
by this gradient plays a crucial role in reviving shock expansion
(\citealt{bhf:95,herant:94}; see \citealt{couch:15a} for the role of
turbulence). 2D simulations also showed that a non-spherical
instabilitiy of the standing accretion shock (SASI;
\citealt{blondin:03,foglizzo:06}) can also help revive the shock.

\begin{figure}[b]
\centering
\includegraphics[width=\columnwidth]{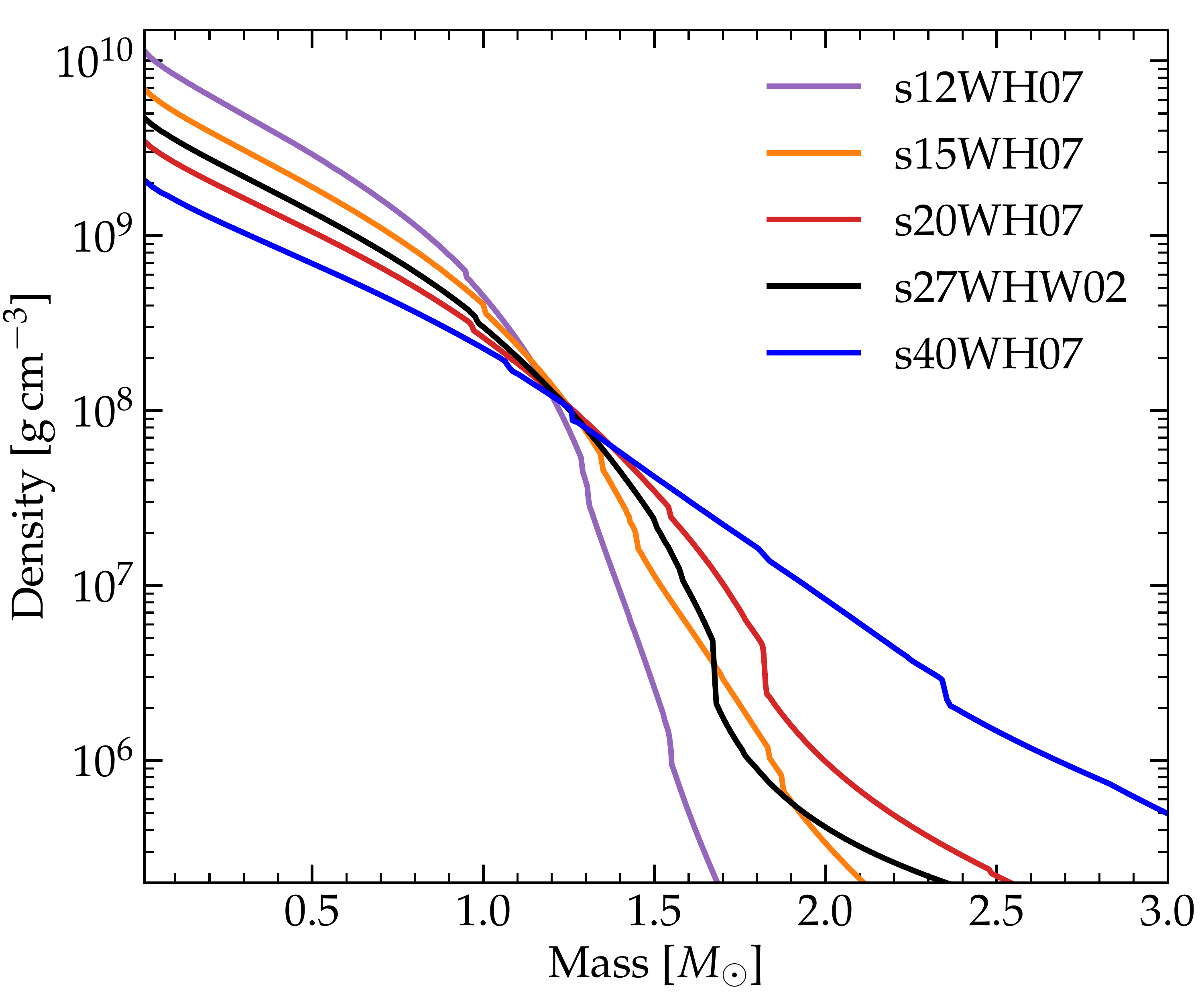}
\caption{The density as a
  function of enclosed mass coordinate for our set of progenitor stars. The
  density profile is the single most important progenitor property
  since it sets the postbounce accretion rate. Note that the
  structures inside $\sim$$1.3-1.4\,M_\odot$ obey a homology
  relationship due to the universal nature of degenerate self-gravitating objects.}
\label{fig:progs}
\end{figure}

The recent availability of petascale supercomputers has enabled the first
detailed 3D CCSN simulations \citep{hanke:13, tamborra:14a, lentz:15,
melson:15a, melson:15b, roberts:16c, takiwaki:14a, takiwaki:16, mueller:17,
summa:17, chan:17}. Comparisons with 2D simulations have shown that 3D is
essential for understanding CCSNe, their explosion mechanism, and for predicting
their multi-messenger observables (cf.~\citealt{couch:15a,janka:16a}).

\smallskip
In this \emph{Letter}, we present a first study of the
progenitor-star dependence of neutrino-driven CCSNe in 3D, covering
zero-age main-sequence masses from $12\,M_\odot$ to $40\,M_\odot$. All
progenitors, except for the $12 \, M_\odot$ star, see shock
runaway by 500 ms after bounce, but in remarkably distinct ways,
depending sensitively on their precollapse structure.

\section{Methods and Setup}

We draw 1D progenitors of $12$, $15$, $20$, and $40\,M_\odot$ from the
set of \cite{woosley:07} (WH07). In addition, we use the $27\,M_\odot$
model of \cite{whw:02} (WHW02) that was simulated in 3D by
\cite{hanke:13} and \cite{roberts:16c}. We plot the progenitor density
profiles in Figure~\ref{fig:progs}.

We simulate core collapse in 1D using \code{GR1D}
\citep{oconnor:13,oconnor:15a} and map to 3D at $20\,\mathrm{ms}$
after bounce for all WH07 progenitors.  The $27\,M_\odot$ model is
mapped at $38\,\mathrm{ms}$ after bounce due to a transposed-digits
error of the lead author. We carry out the 3D simulations with the
open-source 3D general-relativistic (GR) multi-group
radiation-hydrodynamics CCSN code \texttt{Zelmani}
\citep{roberts:16c}. It is based on the \texttt{Einstein Toolkit}
\citep{et:12,moesta:14a}. Neutrino transport is handled in the GR M1
multi-group approximation \citep{shibata:11}. We use three neutrinos
species ($\nu_e$, $\bar{\nu}_e$, and $\nu_x =
[\nu_\mu,\bar{\nu}_\mu,\nu_\tau,\bar{\nu}_\tau]$), and 12 energy
groups, spaced logarithmically with bin-centers between
$1\,\mathrm{MeV}$ and $248\,\mathrm{MeV}$. We employ the subset of
\cite{bruenn:85} neutrino opacities used in \cite{oconnor:13}, but in
3D leave out velocity dependence and inelastic scattering processes
(they are included in 1D). All simulations employ the SFHo equation of
state, which is tuned to fit astrophysical and experimental
constraints \citep{steiner:13b}.
      
The 3D simulations use 8 levels of Cartesian adaptive mesh refinement,
resolving the PNS with $370\,\mathrm{m}$ resolution and the postshock
region with $1.5\,\mathrm{km}$ before shock expansion (see
\citealt{abdikamalov:15,roberts:16c} for resolution studies with our
code).  After the shock has expanded to radii
$\gtrsim$$300\,\mathrm{km}$, we regrid to $3\,\mathrm{km}$ resolution
for the shocked region.  Table~\ref{tab:summary} summarizes key model
properties. All times in this letter are measured relative to core
bounce of each model.

\begin{deluxetable}{lcccc}
  \tablecaption{Model Summary \label{tab:summary}}
  \tablecomments{$\xi_{1.75}$ is the core compactness
    \citep{oconnor:11} measured at bounce at a mass coordinate of
    $1.75\,M_\odot$. $M_\mathrm{ic,b}$ is the mass of the homologous
    core at bounce. $t_\mathrm{f} - t_\mathrm{b}$ is the final
    simulation time relative to core bounce and $\langle
    R_\mathrm{shock,f} \rangle$ is the final average shock
    radius.}
  \tablecolumns{5}
  \tablewidth{\columnwidth}
  \tablehead{
    \colhead{Model}
    &\colhead{$\xi_{1.75}$}
    &\colhead{$M_\mathrm{ic,b}\, [M_\odot]$}
    &\colhead{$t_\mathrm{f} - t_\mathrm{b}\,[\mathrm{ms}]$}
    &\colhead{$\langle R_\mathrm{shock,f} \rangle\,[\mathrm{km}]$}
    }
    \startdata
    s12WH07  & 0.235 & 0.583 &527&123\\
    s15WH07  & 0.580 & 0.576 &597&526\\
    s20WH07  & 0.944 & 0.577 &384&523\\
    s27WHW02 & 0.783 & 0.573 &392&482\\
    s40WH07  & 1.328 & 0.562 &323&614
    \enddata
\end{deluxetable}

\begin{figure*}[t]
\centering
\includegraphics[width=\textwidth]{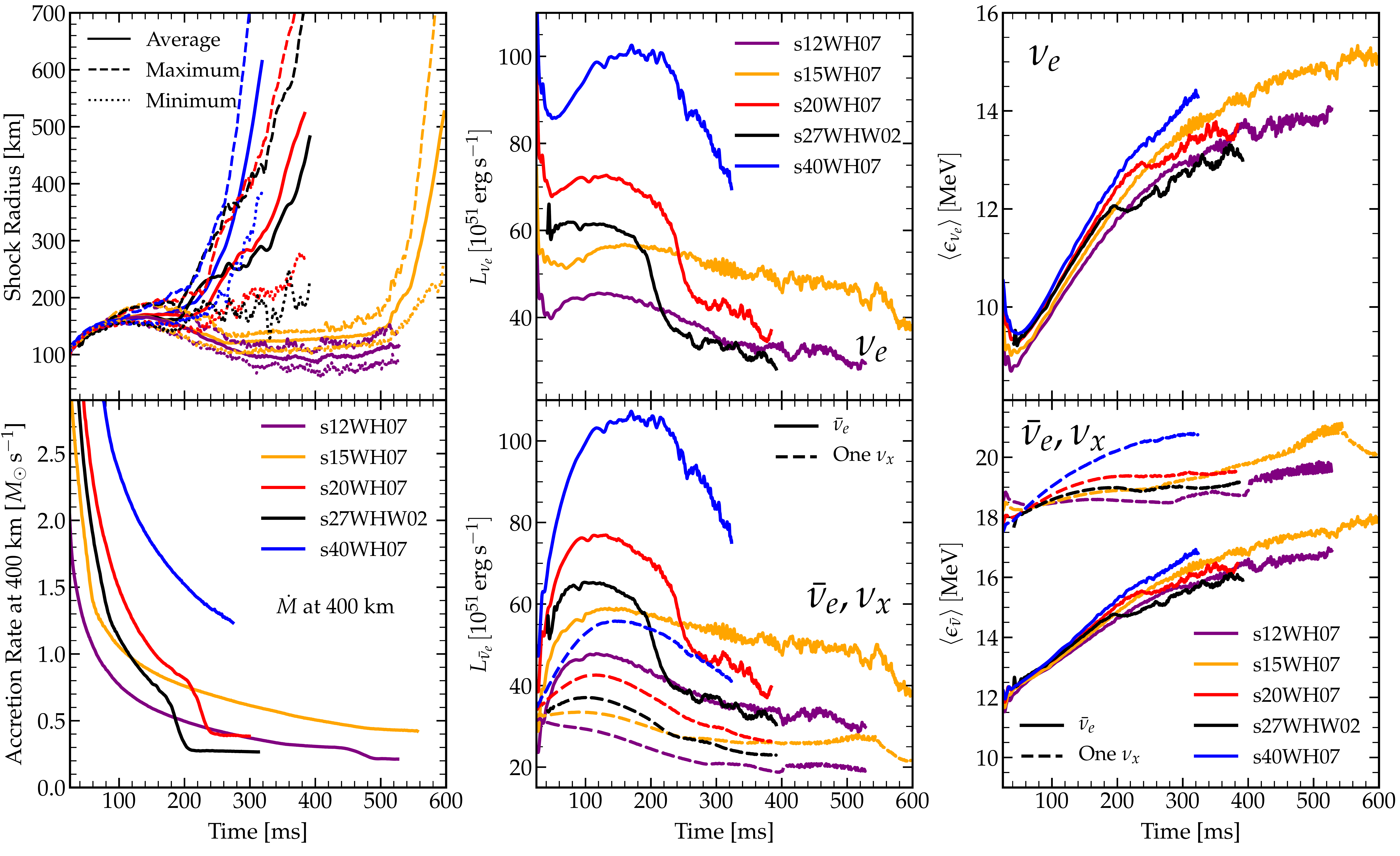}
\caption{Basic radiation hydrodynamics results as a function of time
  after core bounce. The left panel pair depict shock radius (top) and
  accretion rate $\dot{M}$ at $400\,\mathrm{km}$ (bottom). The
  $\dot{M}$ curves terminate when the shock first exceeds that
  radius. The center panels show the electron neutrino luminosities
  (top) and electron antineutrino and heavy-lepton neutrino
  luminosities (bottom), extracted at $450\,\mathrm{km}$. In the right
  panels, we plot the mean electron neutrino (top) and electron
  antineutrino and heavy-lepton neutrino (bottom) energies. Note the
  strong dependence of the $\nu_e$ and $\bar{\nu}_e$ luminosities on
  the accretion rate. The mean energies exhibit much less $\dot{M}$
  sensitivity and their overall increase is driven by the contraction
  of the PNS (cf.~Fig.~\ref{fig:panel2}). Shock runaway occurs in
  s20WH07 and s27WHW02 when the Si/Si-O interface reaches the shock
  and $\dot{M}$ drops. No such drop is necessary to revive s40WH07's
  shock. Model s15WH07 begins shock expansion only after
  $\sim$$500\,\mathrm{ms}$ and model s12WH07 does not experience shock
  runaway by the end of its simulation.}
\label{fig:panel1}
\end{figure*}

\section{Results}

Core bounce occurs when the inner core reaches nuclear density and the
repulsive short-range component of the nuclear force stabilizes its
collapse. With little variation between progenitors, the CCSN shock is
launched from a mass coordinate of $0.56-0.58\,M_\odot$
(Table~\ref{tab:summary}; this is expected, see, e.g.,
\citealt{janka:12b}). The shock first expands rapidly, but quickly
weakens due to the dissociation of heavy nuclei in accreting outer
core material and neutrino losses. It succumbs to the accretion ram
pressure and stalls at a radius of $\sim$$150\,\mathrm{km}$. At this
point, differences in progenitor structure 
begin to matter.

In the bottom-left panel of Fig.~\ref{fig:panel1}, we plot the time
evolution of the mass accretion rate $\dot{M}$ in all
progenitors. Within tens of milliseconds of bounce, the entire iron
core has accreted in all models. Comparing with Fig.~\ref{fig:progs},
we see the subsequent $\dot{M}$ is determined by the progenitor
density profile in the overlying Si and Si-O shells at mass coordinate
$M \gtrsim 1.3-1.9\,M_\odot$.

\begin{figure*}[t]
\centering
\includegraphics[width=\textwidth]{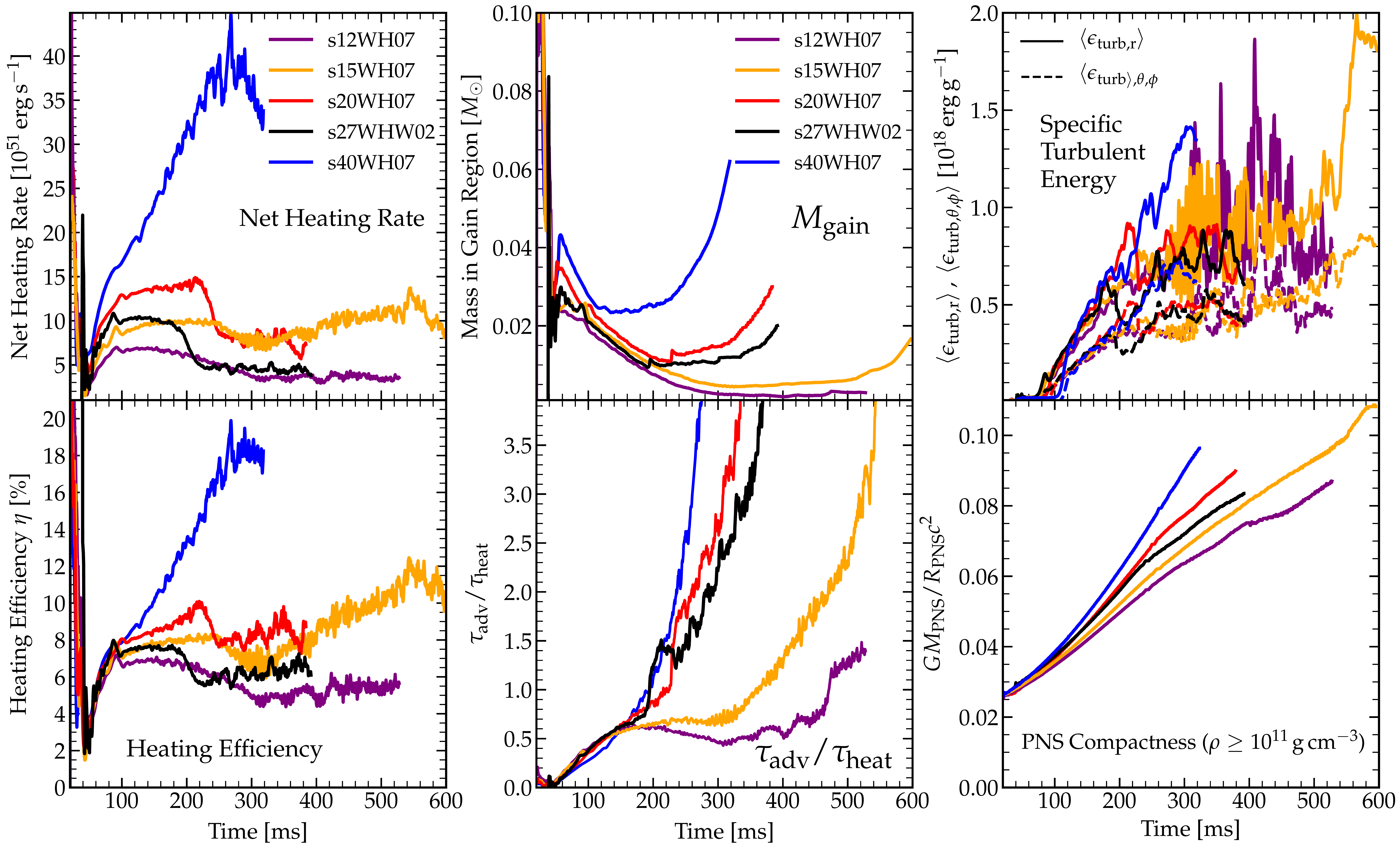}
\caption{Neutrino heating, turbulence, and explosion diagnostics. The
  top-left panel shows the integrated net heating (heating minus
  cooling) rate in the gain region. We plot the heating efficiency
  $\eta = \dot{Q}_\mathrm{net} (L_{\nu_e} + L_{\bar{\nu}_e})^{-1}$ in
  the bottom-left panel. The center panels show the mass contained in
  the gain region (top) and the ratio of the advection timescale
  $\tau_\mathrm{adv}$ to heating timescale $\tau_\mathrm{heat}$
  (bottom). For this, we follow the definition in
  \cite{mueller:12a}. If the ratio is greater than one, conditions are
  favorable for explosion. The top-right panel shows the radial and
  nonradial specific turbulent kinetic energy following the definition
  of \cite{mueller:17} and we plot the PNS compactness in the
  bottom-right panel. Key observations are: (1)
    The onset of shock expansion is preceeded by a stabilization and
    then increase of $M_\mathrm{gain}$. Shock expansion coincides with
    a rapid increase in $\tau_\mathrm{adv}/\tau_\mathrm{heat}$.  (2)
    Model s40WH07 stands out in neutrino heating, due largely to its
    efficient combination of high neutrino luminosity with
    neutrino-driven turbulence that allows it to keep a large amount
    of mass in its gain region.}
\label{fig:panel2}
\end{figure*}

One expects the accrection rate to have multiple important, in part
counteracting roles. First, it sets the accretion ram pressure
$P_\mathrm{ram} = \rho v^2 \propto \dot M M_{\rm PNS}^{1/2}
r_s^{-5/2}$ (the second relationship results from assuming the
accreted material is in free-fall from a large radius), which keeps
the stalled shock from expanding. Second, it regulates the accretion
luminosity $L_\mathrm{acc} \propto \dot{M}
M_\mathrm{PNS}R^{-1}_\mathrm{PNS}$, which is the dominant source of
$\nu_e$ and $\bar{\nu}_e$ luminosity providing energy to the
shock. The hierarchy of $\nu_e$ and $\bar{\nu}_e$ luminosities between
the different progenitor models shown in the center panels of
Fig.~\ref{fig:panel1} directly reflects the $\dot{M}$ order. Finally,
the integrated accretion rate determines mass and radius evolution of
the PNS. To order of magnitude, the mean neutrino energies are
proportional to the temperature at the surface of the PNS. A virial
argument suggests $\langle \epsilon_{\nu_e} \rangle \propto (M_{\rm
  PNS} R^{-1}_{\rm PNS})^\alpha$ (and we find $\alpha = 0.35$ to work
best), which will depend on the integrated accretion rate. At early
times, $M_{\rm PNS} R^{-1}_{\rm PNS}$ is similar among all progenitors
due to the universal structure of the collapsed cores
(Fig.~\ref{fig:panel2}) and therefore the mean neutrino energies are
similar (Fig.~\ref{fig:panel1}), but they start to deviate with time
due to differing accretion histories. Because of all that, one expects
the accretion history to be a determining factor in the CCSN evolution
of a given progenitor.

The impact of progenitor structure on the shock evolution toward
explosion is depicted by the top-left panel of
Fig.~\ref{fig:panel1}. We see three qualitatively different
evolution modes:

\begin{enumerate}[label=(\arabic*),leftmargin=1.5em,topsep=0.2em,itemsep=0.2em]

\item The $40\,M_\odot$ progenitor has the highest postbounce
  $\dot{M}$, translating to the highest neutrino luminosities.  Its
  density profile (Fig.~\ref{fig:progs}) is shallow and smooth and
  there are no quick drops in $\dot{M}$. This model's shock begins to
  deviate substantially from spherical symmetry at
  $\sim$$100\,\mathrm{ms}$ after bounce and shock runaway ensues at
  around $200\,\mathrm{ms}$.  
  
\item The $20\,M_\odot$ and $27\,M_\odot$ models have lower postbounce
  $\dot{M}$, but their density profiles have a steep discontinuity at
  the Si/Si-O shell interface\footnote{The magnitude of the density jump is set
  by the scale of the jump in specific entropy between shells,
  e.g.~\cite{sukhbold:17,suwa:16}.} (cf.~Fig.~\ref{fig:progs}). In both
  models, it is the drop in ram pressure due to the rapidly decreasing $\dot{M}$
  that triggers shock runaway $\sim$$170-200\,\mathrm{ms}$ after bounce. 
     
\item In the $12\,M_\odot$ and $15\,M_\odot$ models with their
  moderate $\dot{M}$ and  low $L_{\nu}$, the shock
  recedes to radii around $100\,\mathrm{km}$. The accretion rate
  gradually decreases, and so do the $\nu_e$ and $\bar{\nu}_e$
  luminosities (central panels of Fig.~\ref{fig:panel1}), while the
  mean neutrino energies increase due to the increasing compactness of
  the PNS (bottom-right panel of Fig.~\ref{fig:panel2}).  Both models
  experiences SASI. Eventually, more than $500\,\mathrm{ms}$ after
  bounce, shock runaway occurs in the $15\, M_\odot$ model.
  The $12\, M_\odot$ model does not experience shock runaway by the end of our
  simulation, but it still has the potential to resume expansion at a later time.
  
\end{enumerate}
\cite{oconnor:15b} and \cite{summa:16} found similar evolutions to
modes (2) and (3) in 2D simulations.

In Fig.~\ref{fig:panel2}, we present diagnostics that help understand
the three evolution modes. Shock expansion is facilitated by increases
in thermal and turbulent pressure that offset the accretion ram
pressure (e.g., \citealt{couch:15a}). Stronger neutrino heating means
more thermal pressure and stronger driving of turbulence. The neutrino
heating rate scales approximately as $ \dot{Q}_\mathrm{heat} \propto
(\langle \epsilon_{\nu_e}^2 \rangle L_{\nu_e} + \langle
\epsilon_{\bar{\nu}_e}^2 \rangle L_{\bar{\nu}_e})R_\mathrm{gain}^{-2}
M_\mathrm{gain} \,\,, $ where $M_\mathrm{gain}$ and $R_\mathrm{gain}$
are the mass contained in the gain region and the gain radius,
respectively. Therefore, the hierarchy of heating rates among the
models mirrors their luminosity hierarchy
(Fig.~\ref{fig:panel2}). Assuming that the majority of the $\nu_e$ and
$\bar \nu_e$ luminosity is powered by accretion, one finds $\dot
Q_{\rm heat} \propto \dot M (M_{\rm PNS} R^{-1}_{\rm PNS})^{1+2\alpha} R_{\rm
  gain}^{-2} M_{\rm gain}$, which implies greater heating for a higher
accretion rate and a more compact PNS for a fixed gain region size and
mass. Interestingly, at early times ($\lesssim$
$80-100\,\mathrm{ms}$), the heating efficiency $\eta =
\dot{Q}_\mathrm{net} (L_{\nu_e} + L_{\bar{\nu}_e})^{-1}$ (where
$\dot{Q}_\mathrm{net}$ is the net heating rate; heating minus cooling)
is independent of progenitor. Since the mean neutrino energies are
very similar at early times, this implies that $M_{\rm gain} R^{-2}_{\rm
  gain}$ is similar for all of the models even though they have
different masses in the gain region (see the top-center panel of
Fig.~\ref{fig:panel2}).

\begin{figure*}[t]
  \centering
\includegraphics[width=0.3\textwidth]{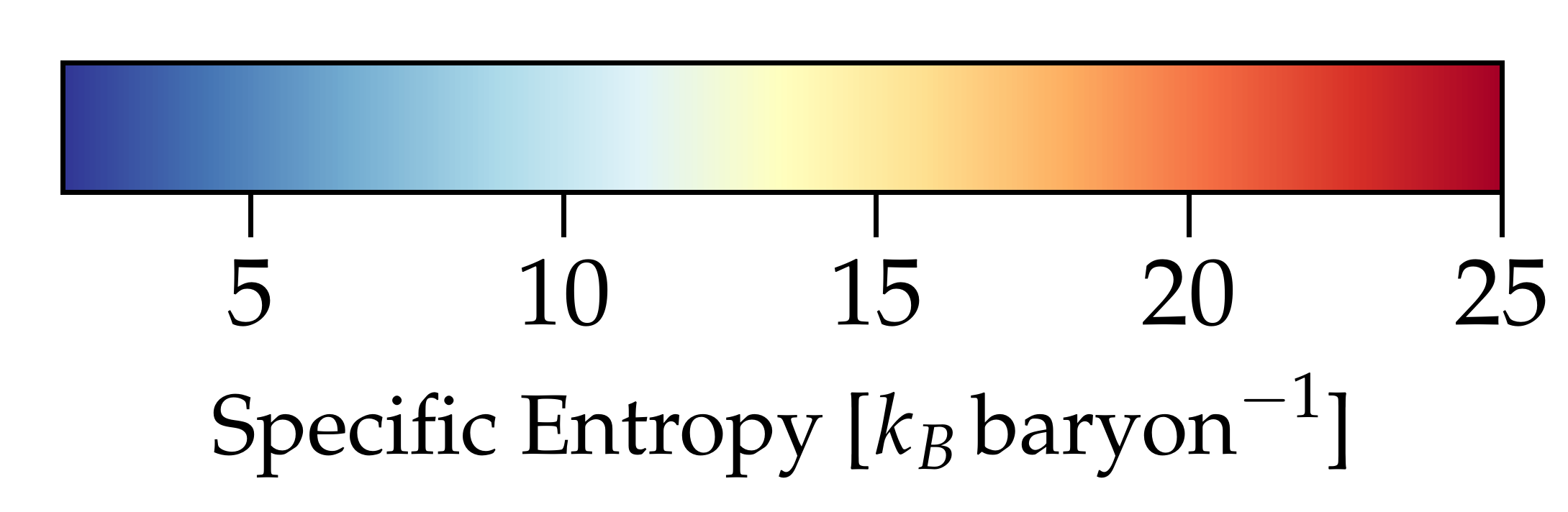}
\includegraphics[width=\textwidth]{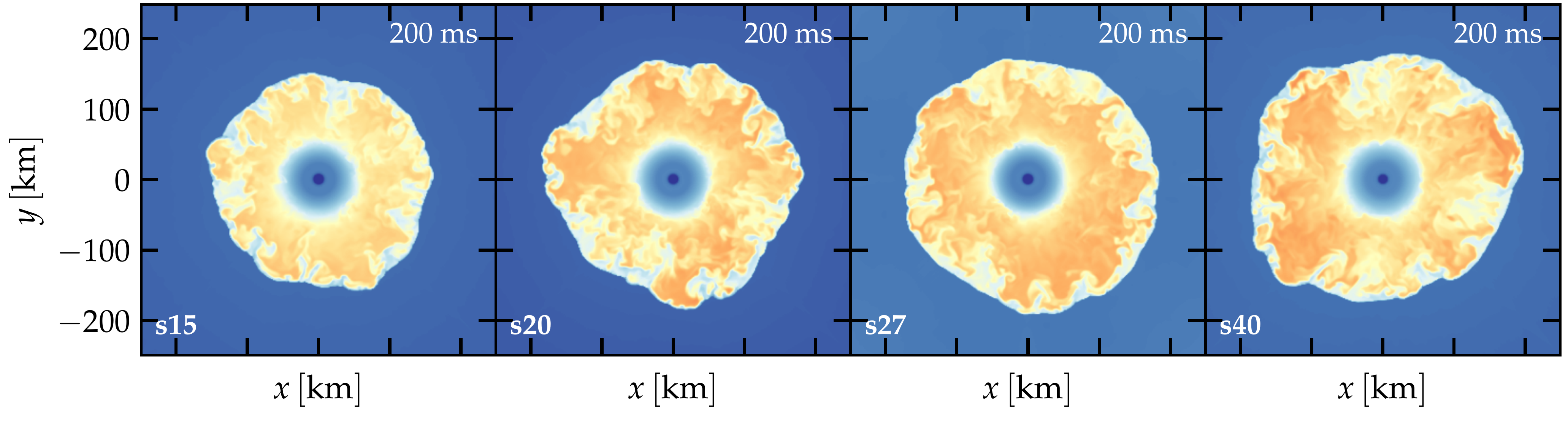}
\includegraphics[width=\textwidth]{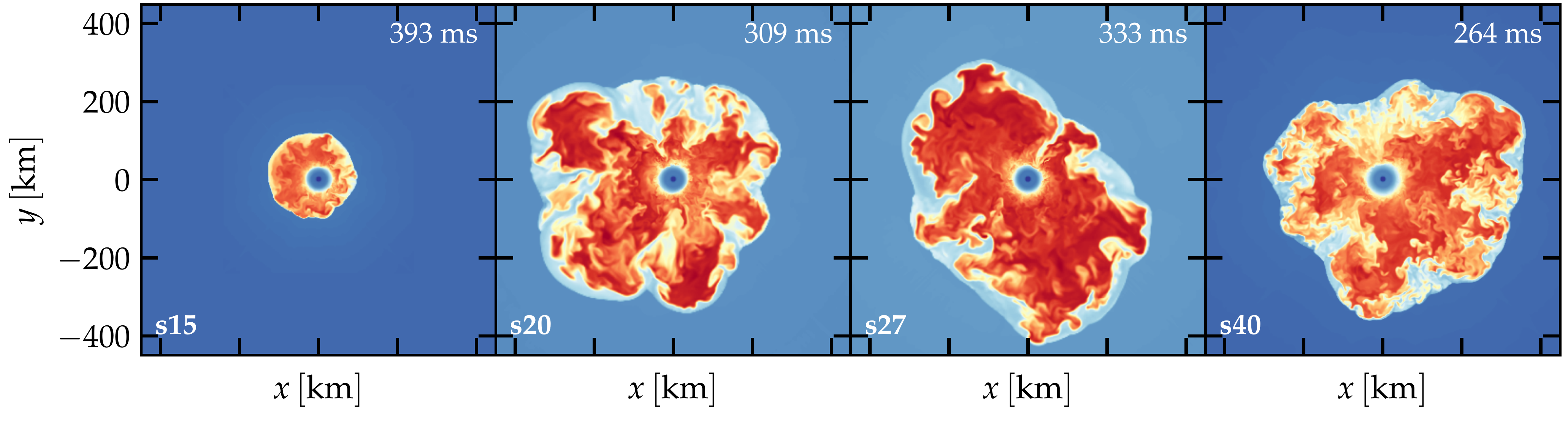}
\includegraphics[width=\textwidth]{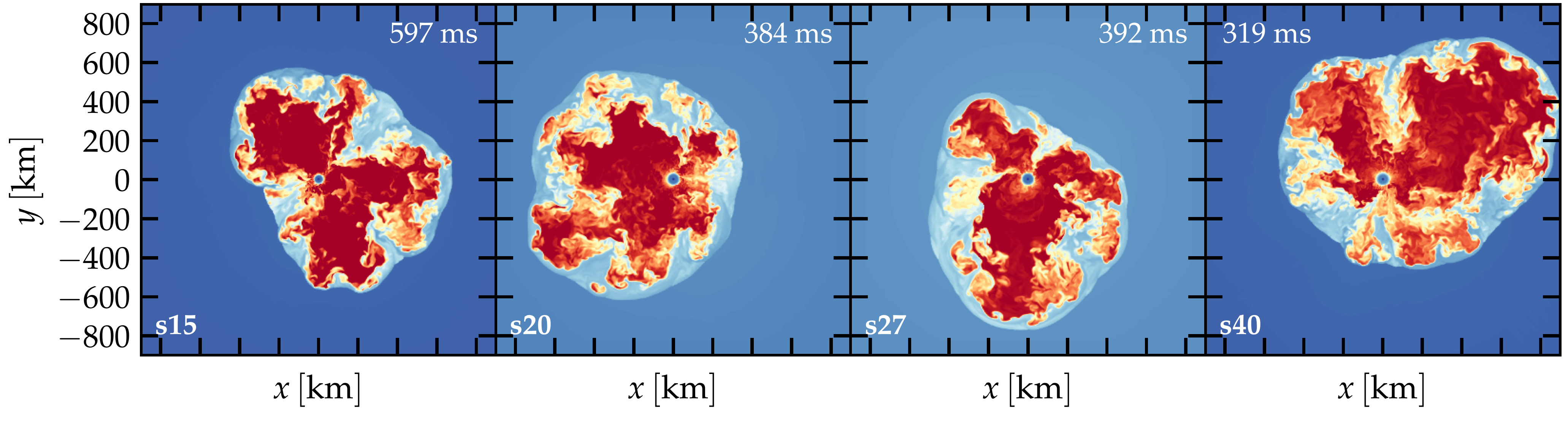}
\includegraphics[width=\textwidth]{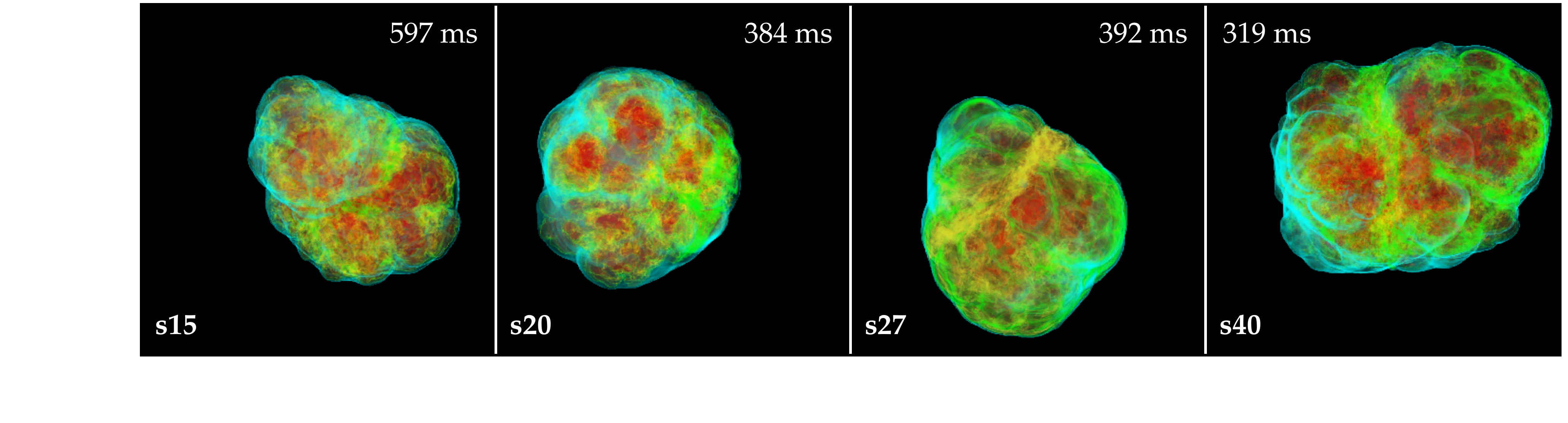}
\caption{Snapshots of the specific entropy in the $x-y$ plane at
  different times in models s15WH07 (first column), s20WH07 (second
  column), s27WHW02 (third column), and s40WH07 (last column). The
  region shown varies between rows with a fixed color range. The top
  row shows all models at $200\,\mathrm{ms}$. Convection is fully
  developed and large-scale asymmetry is beginning to emerge. The
  snapshots of s20WH07, s27WHW02, and s40WH07 in the second row are
  taken at the times when their average shock radii reach
  $300\,\mathrm{km}$. Shock expansion is strongly aspherical. In the
  same row, we show s15WH07 at $393\,\mathrm{ms}$ with a heavily
  SASI-deformed shocked region. The third row shows all models near
  the ends of their simulations. The shock has reached $\gtrsim
  500\,\mathrm{km}$ on average and is running away quickly. The bottom
  row shows 3D entropy volume renderings of the same snapshots.
  The low entropy (\textasciitilde $4-5\, k_B \, \textrm{baryon}^{-1}$) shock
  front is shown in blue, higher entropy regions (\textasciitilde $9
  \, k_B \, \textrm{baryon}^{-1}$ and \textasciitilde $12\, k_B \,
  \textrm{baryon}^{-1}$) are shown in green and yellow, respectively,
  and the highest entropy regions ($\gtrsim$$20-25\, k_B \,
  \textrm{baryon}^{-1}$) are shown in red. }
\label{fig:slices}
\end{figure*}

Neutrino-driven convection begins to grow at
$\sim$$80-100\,\mathrm{ms}$ in all models, as can be seen from the
top-right panel of Fig.~\ref{fig:panel2}, showing the radial and
nonradial specific turbulent kinetic energy in the gain region. It is
fully developed at $200\,\mathrm{ms}$ (Fig.~\ref{fig:slices}).

The specific turbulent energy in the gain region is very similar in
all models and grows with time, although convection starts somewhat
later for s40WH07. When turbulent convection begins in each model, it
briefly reverses the decline of $M_{\rm gain}$. In most models the
decline in $M_{\rm gain}$ continues, but in s40WH07, the onset of
neutrino-driven turbulent convection stabilizes $M_{\rm gain}$, as
shown by the top-center panel of Fig.~\ref{fig:panel2}.  At constant
$M_\mathrm{gain}$ and near constant luminosity, the increase in
neutrino heating in s40WH07 (top-left panel of Fig.~\ref{fig:panel2})
follows from the steep postbounce increase of the neutrino energies
shown in Fig.~\ref{fig:panel1}.

The onset of neutrino-driven convection in model s40WH07 around
$100\,\mathrm{ms}$ after bounce is also reflected in the departure of
its shock from spherical symmetry and the expansion of its maximum
shock radius (Fig.~\ref{fig:panel1}). Conditions for global shock
runaway become gradually more favorable. This can be seen by comparing
the timescale $\tau_\mathrm{adv} \approx M_\mathrm{gain} \dot{M}^{-1} $
it takes for material to advect through the gain layer with
$\tau_\mathrm{heat} \approx |E_\mathrm{gain}|  \dot{Q}^{-1}_\mathrm{net}$,
the timescale for neutrino heating (\citealt{janka:01,thompson:05}; we
follow \citealt{mueller:12a} for implementation details). If
$\tau_\mathrm{adv} / \tau_\mathrm{heat} \gtrsim 1$, it is said that
conditions favor shock runaway. From the bottom-center panel of
Fig.~\ref{fig:panel2}, we see that s40WH07 crosses $\tau_\mathrm{adv}
/ \tau_\mathrm{heat} = 1$ at $\sim$$200\,\mathrm{ms}$ after
bounce. This is at about the time its shock runs away globally.
However, $\tau_\mathrm{adv} / \tau_\mathrm{heat} \gtrsim 1$ seems to
be more of a diagnostic rather than a critical condition for
explosion. Instead, what appears to set model s40WH07 on a positive
track toward shock runaway is the stabilization of the mass in its
gain layer. This occurs when neutrino-driven turbulence sets in, about
$100\,\mathrm{ms}$ before s40WH07 reaches $\tau_\mathrm{adv} /
\tau_\mathrm{heat} \sim 1$.

\begin{figure*}[t]
  \centering
\includegraphics[width=\textwidth]{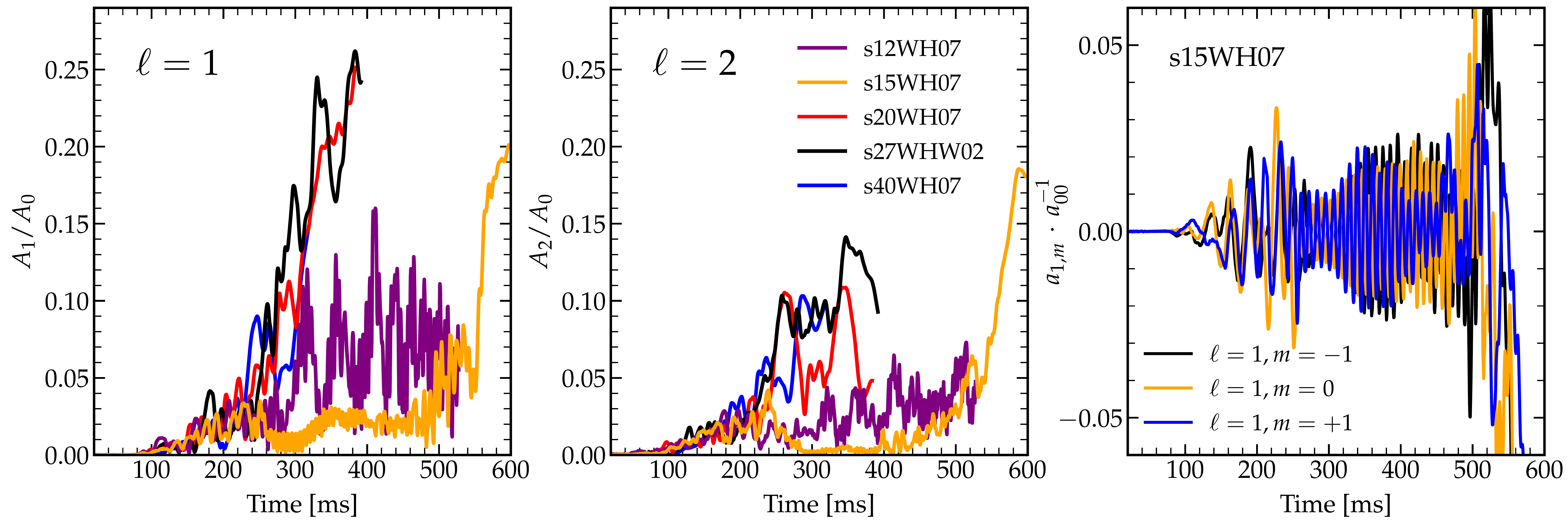}  
\caption{Shock morphology diagnostics. In the left
  panel, we show the normalized rms sum of the $\ell = 1$ coefficient
  of a spherical harmonics decomposition of the shock front. The
  center panel shows the same for the $\ell = 2$ coefficients. In the
  right panel, we present the $\ell = 1, m = \{-1,0,1\}$ mode
  amplitudes for model s15WH07 that attest to its SASI oscillations.
}
\label{fig:panel3}
\end{figure*}

The diagnostics for models s12WH07, s15WH07, s20WH07, and s27WHW02
shown in Fig.~\ref{fig:panel2} are qualitatively very similar until
$\sim$$150-180\,\mathrm{ms}$ after bounce. Neutrino heating scales
roughly with the luminosities, which in turn scale with the accretion
rates (Fig.~\ref{fig:panel1}). $M_\mathrm{gain}$ decreases with the
accretion rate and the heating efficiencies show only moderate
dependence on progenitor model. 

Models that fall into mode (2) defined above, s20WH07 and s27WHW02,
depart from this trend when their Si/Si-O interface reaches the
shock. In s27WHW02, this occurs at $\sim$$180\,\mathrm{ms}$, leading
to the steep drop in $\dot{M}$ seen in the bottom-left panel of
Fig.~\ref{fig:panel1}. Correspondingly, the ram pressure lid on the
shock is lifted, $M_\mathrm{gain}$ stabilizes,
$\tau_\mathrm{adv}/\tau_\mathrm{heat}$ (Fig.~\ref{fig:panel2},
bottom-center) jumps above $1$, and global shock expansion ensues. An
important aspect is that for one advection time of $\tau_\mathrm{adv}
\sim 10-15\,\mathrm{ms}$, the neutrino luminosity and heating rate
remain approximately constant. They drop only once the lower
$\dot{M}$, due now to both the density drop and the shock expansion,
has propagated to the PNS edge. This helps power shock expansion. A
drop of $\dot{M}$ at near-constant luminosity was identified, e.g., by
\cite{suwa:16} as conducive to explosion.  Model s20WH07 mirrors what
we find for s27WHW02.

Models s12WH07 and s15WH07 also mirror each other for most of
their evolution (see Fig.~\ref{fig:panel2}).
Their accretion rates are moderate and neutrino heating and, consequently,
neutrino-driven turbulence are not strong enough to keep the shock on a positive
trajectory. In the following, we focus on s15WH07.

After $\sim$$150\,\mathrm{ms}$, the shock begins to recede despite
near constant heating rates. $\tau_\mathrm{adv}/\tau_\mathrm{heat}$
hovers around $0.5$. The average shock radius settles around
$100\,\mathrm{km}$ and remains there for hundreds of milliseconds, as
shown by the top-left panel of Fig.~\ref{fig:panel1}. We note
that the shock departs substantially from spherical symmetry with
minimum and maximum shock radii differing by $\sim$$40\,\mathrm{km}$
in s15WH07 at $200-500\,\mathrm{ms}$ after bounce.

Shock recession leads to short advection times and conditions
favorable for the growth of SASI (e.g.,
\citealt{scheck:08,fernandez:15a}).  SASI with a substantial spiral
mode \citep{fernandez:10} develops in s15WH07 and s12WH07. In
Fig.~\ref{fig:slices} (second row, first column), we show a snapshot
of the lopsided specific entropy distribution in the $x-y$ plane of
model s15WH07 at $393\,\mathrm{ms}$ after bounce, which shows
a SASI-deformed shock.  In the right panel of
Fig.~\ref{fig:panel3}, we plot the normalized $\ell = 1, m =
\{-1,0,1\}$ mode amplitudes of a spherical harmonics expansion of the
shock front. The mode amplitudes grow $100-200\,\mathrm{ms}$ after
bounce when neutrino-driven convection is present (cf.~top and second
row, first column of Fig.~\ref{fig:slices}). They become clearly
oscillatory and persistent once SASI develops,
$\sim$$250\,\mathrm{ms}$ after bounce.

With the help of SASI and a growing specific turbulent energy in the
gain region, neutrino heating gradually brings s15WH07 back onto a
positive track toward shock revival. Figure~\ref{fig:panel2} shows
that the mass $M_\mathrm{gain}$ contained in the gain region
stabilizes around $300\,\mathrm{ms}$ after bounce. While the neutrino
luminosity continues its slow decline (Fig.~\ref{fig:panel1}), heating
rate and heating efficiency both increase since $M_\mathrm{gain}$ is
stabilized and neutrino energies increase. The timescales ratio
$\tau_\mathrm{adv} / \tau_\mathrm{heat}$ follows and slowly reaches
unity around $400\,\mathrm{ms}$ after bounce. Yet, shock expansion
does not follow immediately and s15WH07 straddles the threshold of
shock runaway for another $\sim$$100\,\mathrm{ms}$ until heating and
turbulence are strong enough to overcome the slowly decreasing ram
pressure.

In contrast to s15WH07, model s12WH07 does not experience shock
runaway within the simulation time. However, at the end of the
simulation, s12WH07 has $\tau_{\rm adv}/\tau_{\rm heat} > 1$ and the
mass in the gain region has stabilized, both of which suggest it is on
the path to shock revival.

Once the average shock radius reaches $\sim$$300\,\mathrm{km}$ shock
expansion rapidly accelerates in all models. In the third
row of Fig.~\ref{fig:slices}, we show $x-y$ slices of the specific
entropy at the time the average shock radius reaches
$300\,\mathrm{km}$ in models s20WH07, s27WHW02, and s40WH07. The
expanding shock already exhibits large asymmetry in these models,
which only grows as the expansion accelerates. This can be appreciate
from the third and forth rows of Fig.~\ref{fig:slices}.

The asymmetry of the expanding shock can be understood more
quantitatively by considering Fig.~\ref{fig:panel3}.  There, we plot
the normalized rms mode amplitudes (summed over all $m$) for $\ell =
1$ (top panel) and $\ell = 2$ (bottom panel). As seen in previous
exploding 3D simulations (e.g.,
\citealt{lentz:15,melson:15b,roberts:16c,mueller:17}), all models
develop a strong, growing, and dominant $\ell = 1$ asymmetry.  There
is also power in $\ell = 2$ and in higher modes (not shown),
reflecting the complex morphology of the expanding shock shown in the
two bottom rows of Fig.~\ref{fig:slices}.

\begin{figure}[t]
  \centering
\includegraphics[width=\columnwidth]{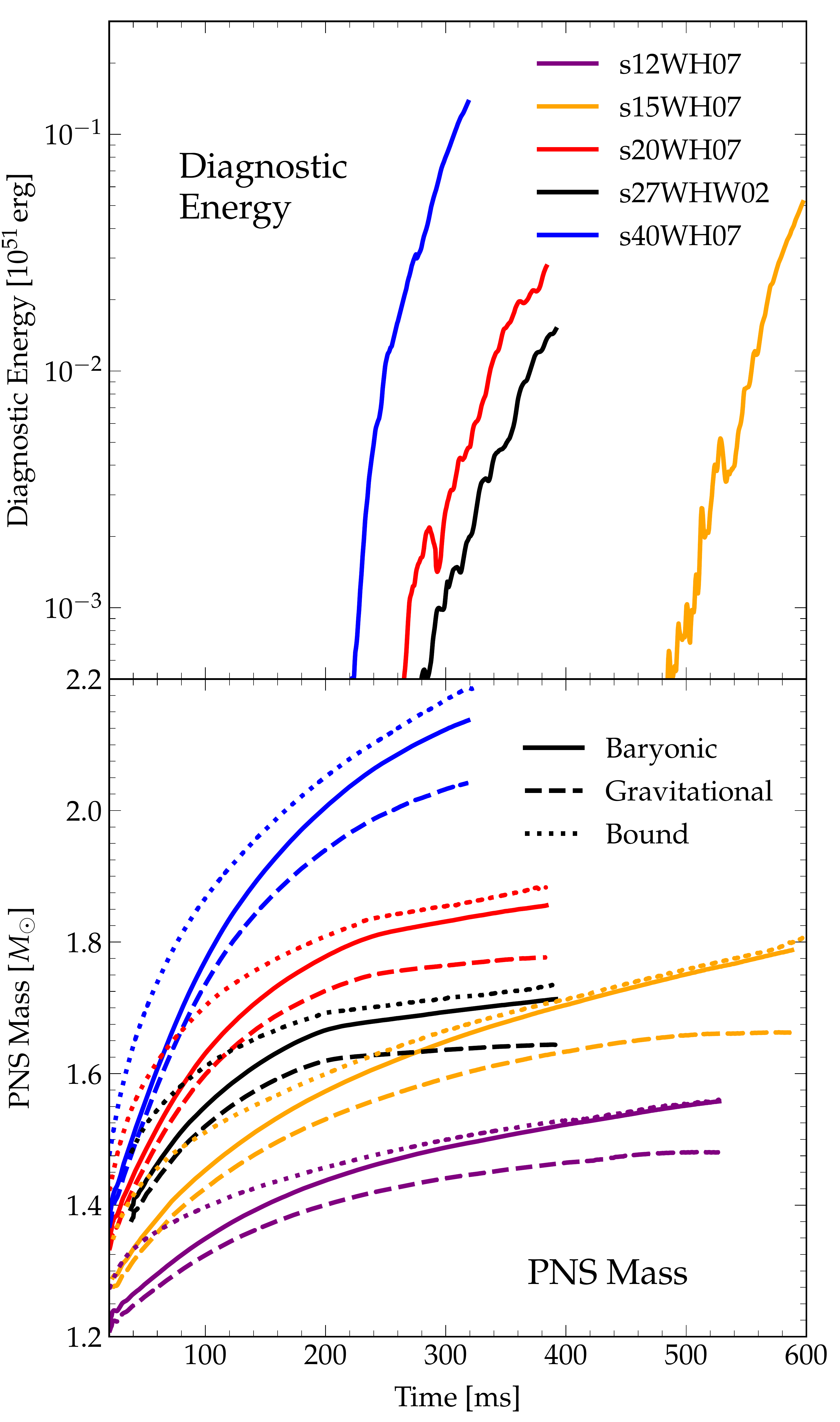}  
\caption{The top panel shows the diagnostic explosion energy (using
  the definition of \citealt{mueller:12a}). It is rapidly growing at
  the end of our simulations. It does not include positive
  contributions from nuclear recombination. The shock still has to
  overcome the binding energy of the overlying stellar material, which
  is greater than the diagnostic energy in all models at the final
  simulated time. In the bottom panel, we plot the baryonic and
  gravitational PNS mass inside the $10^{11}\,\mathrm{g\,cm}^{-3}$
  density contour. Also plotted is the bound baryonic mass inside the
  expanding shock. Model s40WH07's PNS is still accreting at
  $\sim$$0.45\,M_\odot\,s^{-1}$ at the end of the simulation. It will
  thus likely exceed the maximum mass that can be supported by the
  SFHo EOS ($2.06\,M_\odot$ for a cold NS, $\sim$$10-20\%$ more for a
  hot PNS; \citealt{oconnor:11}) and collapse to a black hole.}
\label{fig:panel4}
\end{figure}

In the top panel of Fig.~\ref{fig:panel4}, we plot the diagnostic
energy (Eq.~2 of \citealt{mueller:12a}) of unbound material in our
models. All our models are in the rapid shock expansion phase at the
end of their simulations and the diagnostic energy grows quickly.
This energy does not include possible contributions from nuclear
recombination (up to $1.7\times10^{51}\,\mathrm{erg}$ per
$0.1\,M_\odot$ recombining to $^{56}\mathrm{Fe}$). It also
does not factor in the ``overburden'' \citep{bruenn:16}, the binding
energy of the overlying stellar material. Using the PNS masses in
Fig.~\ref{fig:panel4} as approximate mass cuts, we find, based on
precollapse structures, overburdens of $(17, 7.7, 6, 3.3, 1.3
)\times 10^{50}\,\mathrm{erg}$ for the
$(40, 27, 20, 15)\, M_\odot$ models, respectively. Hence, the
diagnostic energies still must grow before the overall energy budget
becomes positive and typical CCSN energies are reached. The PNS, via
its neutrino-driven wind, will continue to inject energy into the
explosion for seconds. Luminosity from continuing accretion,
facilitated by the highly asymmetric shock expansion, will also
contribute. \cite{bruenn:16} (in 2D) and \cite{mueller:17} (in 3D)
have shown that in this way typical final explosion energies of
$10^{50} - 10^{51}\,\mathrm{erg}$ can be reached by the neutrino
mechanism.

Model s40WH07, however, is a special case. Its PNS has a gravitational
(baryonic) mass of $\sim$$2.05\,M_\odot$ ($\sim$$2.13\,M_\odot$) at
the end of the simulation and it is still accreting at a rate of
$\sim$$0.45\,M_\odot\,\mathrm{s}^{-1}$ due to the asymmetry of the
explosion. The cold-NS maximum gravitational mass supported by the
SFHo EOS is $2.06\,M_\odot$. The hot PNS in our non-exploding 1D
simulation of this progenitor collapses to a black hole at a
gravitational mass of $2.33\,M_\odot$. Hence, unless the accretion
rate drops substantially, model s40WH07 will form a black hole within
$\sim$$600-700\,\mathrm{ms}$ from the end of our simulation. This will
shut off energy injection into the expanding shock, resulting in an
anemic explosion or complete failure \citep{chan:17}. 

The gravitational PNS mass in the other models has more or less
leveled off at the end of the simulation due to the competition of
neutrino cooling with moderate amounts of continuing accretion.  Using
 $E_\mathrm{bind} \approx 0.084 M_\odot c^2
(M_\mathrm{grav}/M_\odot)^2$ \citep{lattimer:01} and the bound
baryonic masses shown in Fig.~\ref{fig:panel4}, we estimate
(lower-limit) final remnant NS masses of $(1.59,1.65,1.54)\,M_\odot$
for s15WH07, s20WH07, and s27WHW02, respectively.

\section{Discussion and Conclusions}

Simple answers are not to be had in 3D CCSN theory.
Our simulations suggest a complicated interplay of accretion rate,
neutrino heating, and 3D fluid dynamics that determines the resulting
CCSN dynamics and final outcome.

The three ``CCSN evolution modes'' we identify
depend on progenitor structure as follows:

(1) Massive cores with high compactness ($\xi_M = (M/M_\odot)
(R[M]/1000\,\mathrm{km})^{-1},
M=1.75-2.5\,M_\odot$;\citealt{oconnor:11}) and without large density
drops at shell interfaces develop early neutrino-driven, turbulence
facilitated shock runaway, but likely make black holes, with or
without the shock exploding the star.

(2) Cores with intermediate compactness and with a substantial density
drop at the Si/Si-O interface develop neutrino-driven, turbulence
facilitated explosions when this interface reaches the shock. They make
relatively massive NSs with $M\gtrsim1.5\,M_\odot$.

(3) Cores with moderate to low compactness and without precipitous
density drop have receding shocks that develop SASI and run away
only at late times once SASI, neutrino heating, and turbulence have
established favorable conditions. Due to the late explosions, the resulting
NSs are also relatively massive ($M\gtrsim1.4-1.5\,M_\odot$).

Modes (2) and (3) were seen previously in the 2D simulations of
\cite{oconnor:15b} and \cite{summa:16}. Mode (1), for the most extreme
progenitors like s40WH07, is new. \cite{pan:17} recently simulated
this progenitor in 2D with the same EOS, but did not find an
explosion. \cite{chan:17} simulated a different $40\,M_\odot$
progenitor in 3D and modified neutrino opacities to obtain an
explosion. In our simulation of s40WH07, turbulence driven by neutrino
heating is essential for creating conditions allowing shock
runaway. The simulation of \cite{pan:17} appears to have much weaker
turbulence. The strength of CCSN turbulence is sensitive not only to
neutrino heating, but also to the magnitude of perturbations that
enter through the shock
\citep[e.g.,][]{couch:13d,mueller:15,mueller:17}.  Hence, the
differences between our simulations and those of others could be due
to the relatively large numerical perturbations imposed by our
Cartesian grid (e.g., \citealt{ott:13a}). This could, perhaps, explain
why we find explosions for s27WHW02 and s20WH07 that did not explode
in the spherical-coordinates 3D simulations of \cite{hanke:13} and
\cite{tamborra:14a}, and \cite{melson:15b}, respectively. Another piece of
evidence for this argument is that our simulations of s20WH07 and
s27WHW02 are at all times closer to explosion than their 2D
counterparts in \cite{summa:16}, despite the fact that other studies
have shown that 2D is more conducive to explosion than 3D due to its
(unphysical) inverse turbulent cascade (e.g.,
\citealt{couch:13b,couch:15a,lentz:15}). However, one should keep in
mind that \cite{hanke:13}, \cite{tamborra:14a}, \cite{melson:15b}, and
\cite{summa:16} used a different EOS, as well as different
approximations to the neutrino transport and different neutrino
opacities.

Presupernova stars in the wild have physical perturbations in
their iron cores, Si, and Si-O shells. Determining their 
properties requires full 3D stellar evolution simulations
of the final phase before core collapse
\citep{couch:15b,mueller:16,cristini:17}.

Our simulations show that the development of large-scale asymmetric
explosions with dominant $\ell = 1$ components is a generic outcome
and independent of progenitor in the mass range considered here. While
we do not investigate them here, NS and black hole birth kicks
require such asymmetric mass ejection (e.g.,
\citealt{mueller:17,janka:13}).

Our three CCSN evolution modes produce black holes or massive NSs
($M\gtrsim1.4-1.5\,M_\odot$). Hence, there must be a fourth CCSN mode
producing lower-mass NSs. Low-mass progenitors with O-Ne and the
lowest-mass iron core progenitors ($M \lesssim 10\,M_\odot$) explode
even in 1D due to their very steep density profiles (e.g.,
\citealt{kitaura:06, melson:15a}).  They could be responsible for
low-mass NSs. They have less mass to eject and their explosions are
also likely to be less asymmetric, leading to low kick velocities.

\smallskip
There are many ingredients to the CCSN phenomenon. Here we
investigated for the first time the progenitor dependence in
3D. Others have recently investigated the role of rotation on the 3D
neutrino mechanism (e.g., \citealt{takiwaki:16,summa:17}). More such
studies are needed to also investigate magnetohydrodynamic effects,
the impact of various neutrino transport approximations (see
\citealt{richers:17b} for recent progress there), differences in
microphysics (EOS, neutrino interactions), and other numerical issues
such as hydrodynamics methods, grid geometries, and resolution.

\medskip
We acknowledge helpful discussions with D.~Radice, H.~Nagakura,
Y.~Suwa, K.~Kiuchi, M.~Shibata, J.~Takeda, H.-T.~Janka, R.~Bollig,
M.~Obergaulinger, S.~Couch, E.~O'Connor, P.~M\"osta, and
K.~S.~Thorne. This research was partially supported by NSF grants
CAREER PHY-1151197, PHY-1404569, OAC-1550514, and AST-1333520, and the
Sherman Fairchild Foundation. The simulations took over a year to
complete and were carried out on $\mathcal{O}(10,000)$ compute cores
of NSF/NCSA Blue Waters (PRAC ACI-1440083, OCI-0725070 and
ACI-1238993), of Edison at the National Energy Research Scientific
Computing Center, a DOE Office of Science User Facility supported by
the Office of Science of the U.S. Department of Energy under Contract
No.\ DE-AC02-05CH11231, and of the Texas Advanced Computing Center
Stampede and Stampede2 clusters under NSF XSEDE allocation
TG-PHY100033. This article has been assigned Yukawa Institute for
Theoretical Physics report No.~YITP-17-122.

\end{document}